\documentclass[twocolumn]{jpsj2}
\usepackage{graphicx,rotating,subfigure,amsmath,amsfonts,amssymb,delarray}
\renewcommand{\vec}[1]{\boldsymbol #1}

\newcommand{\im}{\text{i}}

\makeatletter
\def\oc@movep{\@tempa}\let\@citey\oc@movep
\renewcommand\@cite[1]{[#1]}

\makeatother
\newcommand{\scite}{\cite}

\def\runauthor{Jesko \textsc{Sirker} {\it et al.}}

\title{The field-induced magnetic ordering transition in TlCuCl$_3$} 
\author{Jesko \textsc{Sirker}, Alexander \textsc{Wei{\ss}e} and Oleg P.~\textsc{Sushkov}}
\inst{School of Physics, The University of New South Wales,
  Sydney 2052, Australia}
\recdate{\today} \abst{ In the first part we investigate in detail if the
  Bose-Einstein condensation scenario for magnons can quantitatively explain
  the observed field-induced magnetic ordering in TlCuCl$_3$. We use a
  bond-operator approach to map the spin system onto hard-core bosons and
  exactly account for the hard-core constraint in the dilute limit. We solve
  the hard-core model within the Hartree-Fock-Popov approximation and discuss
  its validity and the consequences of this approximation for the critical
  properties. In the second part the effects of spin-phonon and spin-orbit
  coupling are discussed within this framework. We show that the experimental
  magnetization and specific heat data are well described if a certain type of
  anisotropy is included.  We also present predictions for the quasiparticle
  gap which might be tested in the future.}
\kword{Spin-gap systems, Bose-Einstein condensation, TlCuCl$_3$}
\begin{document}
\maketitle
\section{Introduction}
Antiferromagnetic systems such as integer-spin chains, spin-1/2 ladders with
an even number of legs or systems where the spins form a network of weakly
coupled dimers share the common property that an excitation gap $\Delta$
between the singlet ground state and the triplet state exists. An applied
magnetic field $H$ leads to a Zeeman splitting of the massive triplet with the
lowest mode crossing the ground state at a critical field $H_g=\Delta/g\mu_B$.
The ground state for $H>H_g$ can be regarded as a Bose-Einstein condensate
(BEC) of this bosonic mode \scite{AffleckBEC,GiamarchiTsvelik,NikuniOshikawa}.
This is particularly easy to understand for $S=1/2$ spins where the
bond-operator (BO) representation
\begin{equation}
\label{BO1} 
S^\alpha_{1,2} = \frac{1}{2}(\pm t_\alpha \pm t_\alpha^\dagger 
-\im\vec{\epsilon}_{\alpha\beta\gamma}t_\beta^\dagger t_\gamma)
\end{equation}
of two neighbouring spins $\vec{S}_{1,2}$ in terms of bosonic triplet creation
(annihilation) operators $t_\alpha^\dagger$ ($t_\alpha$) is exact
\scite{Chubukov,SachdevBhatt,KotovSushkov}. Using (\ref{BO1}) the
spin-Hamiltonian
%% in terms of spin operators $\vec{S}$ 
can be mapped exactly onto a Hamiltonian in terms of the bosonic operators
$t_\alpha$. The magnetic field $H$ becomes the chemical potential $\mu$ for
the magnons and the concept of BEC is in principle directly applicable. Note,
however, that the magnons are subject to the hard-core constraint $t_{\beta
  i}^\dagger t_{\alpha i}^\dagger = 0$, i.e., only one triplet is allowed at
the bond $i$. The magnons therefore form an {\it interacting} Bose gas and the
problem in terms of bosonic operators is in general as complicated as that in
terms of the original spin operators. The BEC concept for the field-induced
magnetic ordering in spin-gap systems becomes only helpful if the magnon
density $n$ is small. More precisely, the average distance between the magnons
$l\sim n^{-1/3}$ should be much larger than the s-wave scattering length $a$
which is the characteristic length scale representing the influence of the
repulsive potential. This implies that $a/l\sim n^{1/3}a\ll 1$ so that the
magnons have to be dilute.  In this case the well-established gas
approximation \scite{FetterWalecka} which involves a systematic expansion in
terms of the small parameter $n^{1/3}a$ is applicable and even the finite
temperature properties of the interacting Bose gas can be studied analytically
\scite{GriffinRep}. In TlCuCl$_3$ this situation is realized in magnetic fields
$H\sim 6-7$ T and this compound has therefore been studied extensively in
recent years
\scite{NikuniOshikawa,MatsumotoNormandPRL,MatsumotoNormandPRB,CavadiniHeigold,CavadiniHeigoldCM,ChoiGuentherodt,ShermanLemmens,SirkerWeisse,SirkerWeisse2,DasguptaValenti,MisguichOshikawa,OosawaIshii,GlazkovSmirnov,OosawaKatori}.
Inelastic neutron scattering (INS) \scite{CavadiniHeigold} has revealed that
TlCuCl$_3$ has an excitation gap $\Delta \approx 0.7$~meV in zero magnetic
field and a bandwidth $W\sim 6.3$~meV. The dimers in this compound are formed
by the $S=1/2$ spins of neighbouring Cu$^{2+}$ ions and weaker interdimer
interactions are mediated by the Cl$^-$ ions yielding a three dimensional
dimer network. On the theoretical side it has been shown that the measured
magnetisation curves can be qualitatively reproduced within the BEC picture
\scite{NikuniOshikawa}. In addition, the magnon dispersion has been described
by using the BO formalism \scite{MatsumotoNormandPRL}. More recently we pointed
out that anisotropies induced by spin-orbit coupling can influence BEC in
spin-gap systems dramatically. Taking such anisotropies into account we have
shown that for TlCuCl$_3$ it is possible to obtain good quantitative agreement
between the measured magnetisation curves and those calculated within the BEC
framework \scite{SirkerWeisse}.

In the present paper we want to extend \scite{SirkerWeisse} with respect to the
following aspects: In section \ref{secBO} we will show how the hard-core
constraint can be taken into account for dilute magnons beyond the mean-field
level thus improving the results in Ref.~\scite{MatsumotoNormandPRL}. In
particular, this approach allows us to calculate the magnon-magnon scattering
amplitude directly. In section \ref{HFPA} we will discuss the validity of the
Hartree-Fock-Popov approximation (HFPA) which we use to solve the hard-core
boson model. We will also investigate how this approximation affects the
critical properties of the model.
%% For TlCuCl$_3$ we show that the experimental found ``power law'' $n_c\sim
%% T^\phi$ with $\phi\sim 2$ is in fact no universal power law and depends on the
%% dispersion relation and the considered energy scale. 
In section \ref{secSO} we discuss the influence of crystal-field anisotropies
for the BEC. We also investigate in more detail the case of a staggered $g$
tensor and/or antisymmetric (Dzyaloshinsky-Moriya) spin interactions proposed
in Ref.~\scite{SirkerWeisse} to explain the measured magnetisation curves. In
particular we will show results for different orientations of the magnetic
field. In section \ref{specHeat} we compare the results of our theory for the
specific heat with experimental data. The last section presents a summary and
conclusions.
\section{Bond-operator formalism}
\label{secBO}
The BO technique starts from the strong coupling ground state $|s\rangle$
where each dimer at bond $i$ forms a singlet $|i,s\rangle$. The operators
$t^\dagger_{i\alpha}$ then create local triplet excitations $|i,\alpha\rangle
= t^\dagger_{i\alpha}\left|i,s\rangle\right.$
%% where $\alpha = +,0,-$ is the spiral index.
with $\left|i,+\rangle\right. = -\left|\uparrow\uparrow\rangle\right.$,
$\left|i,-\rangle\right. = \left|\downarrow\downarrow\rangle\right.$ and
$\left|i,0\rangle\right. =
(\left|\uparrow\downarrow\rangle\right.+\left|\downarrow\uparrow\rangle\right.)/\sqrt{2}$.
For TlCuCl$_3$ an effective Heisenberg-type Hamiltonian has been derived in
\scite{CavadiniHeigoldCM,MatsumotoNormandPRL,MatsumotoNormandPRB} containing
the intradimer coupling $J$ and three interdimer couplings $J_a, J_{a2c},
J_{abc}$. In these works it has been argued that the exchange paths
corresponding to these interdimer couplings are most important. This has been
confirmed in a recent electronic structure calculation \scite{DasguptaValenti}.
We have checked that including additional exchange paths which might be also
of some relevance \scite{DasguptaValenti} does not significantly improve the
results presented here. We therefore restrict ourselves to the model
considered in \scite{MatsumotoNormandPRL,MatsumotoNormandPRB}.
%% which makes it also easier to compare our results with theirs.
Using (\ref{BO1}) and retaining only terms bilinear in the triplet operators
one easily derives \scite{MatsumotoNormandPRL}
\begin{equation}
\label{eq1}
H = \sum_{\vec{k}\alpha}\left\{\left(A_{\vec{k}}+\alpha g\mu_B H\right) t^\dagger_{\vec{k}\alpha}
t_{\vec{k}\alpha} +
\frac{B_{\vec{k}}}{2}\left(t_{\vec{k}\alpha}t_{-\vec{k}\bar{\alpha}}+h.c.\right) \right\} 
\end{equation}
with $A_{\vec{k}} = J + f_{\vec{k}} + g_{\vec{k}}$, $B_{\vec{k}} = f_{\vec{k}}
+ g_{\vec{k}}$, $f_{\vec{k}} = J_a \cos k_x + J_{a2c} \cos(2k_x+k_z)$,
$g_{\vec{k}} = 2J_{abc}\cos(k_x+k_z/2)\cos(k_y/2)$, $\bar{\alpha} = -\alpha$
where $-\pi\leq k_x,k_y\leq\pi$ and $-2\pi\leq k_z\leq 2\pi$. By the
Bogoliubov transform $t_{\vec{k}\alpha} =
u_{\vec{k}}\tilde{t}_{\vec{k}\alpha}+v_{\vec{k}}\tilde{t}^\dagger_{-\vec{k},\bar{\alpha}}$
the Hamiltonian gets diagonal %% in the new operators $\tilde{t}$ 
with eigenvalues
\begin{equation}
\label{eq2}
\omega_{\vec{k}\alpha}  = \sqrt{A_{\vec{k}}^2-B_{\vec{k}}^2} -\alpha g\mu_B H
\end{equation}
%% $\omega_{k\alpha}$ and Bogoliubov coefficients $u_k$, $v_k$ which are given by
%% \begin{subequations}
and Bogoliubov coefficients 
\begin{equation}
\label{eq3}
u_{\vec{k}}^2, v_{\vec{k}}^2 = \pm \frac{1}{2} + \frac{A_{\vec{k}}}{2\omega_{\vec{k}0}}\; .
\end{equation}
%% \end{subequations}
A good fit of the measured magnon dispersion \scite{CavadiniHeigold} can be
obtained from (\ref{eq2}) using $J, J_a, J_{a2c}, J_{abc}$ as fitting
parameters. Because these parameters are not very sensitive to the gap value
whereas the magnetic properties %% of the system 
for fields $H\gtrsim H_g$ crucially depend on $\Delta$ we prefer to fix the
gap and use it as a constraint in the fitting procedure. Results for
$\Delta=0.65, 0.8$ and $1.0$~meV are given to the left of each column in table
\ref{tab1} and are consistent with \scite{MatsumotoNormandPRL}.
%% As shown by Matsumoto {\it et al.}~\scite{MatsumotoNormandPRL} an excellent fit
%% of the experimentally observed magnon dispersion \scite{CavadiniHeigold} can be
%% obtained from Eq.~(\ref{eq2}) using $J, J_a, J_{a2c}, J_{abc}$ as fitting
%% parameters. 
%% %% However, as the magnetic properties of the system for fields
%% %% $H\gtrsim H_c$ crucially depend on the magnon dispersion near the band minimum
%% %% we have fixed different gap values in a range consistent with experiment and
%% %% used those values as a constraint in the fitting procedure. The results of the
%% %% fits are given in table \ref{tab1}.
%% However, as the magnetic properties of the system for fields $H\gtrsim H_c$
%% depend very sensitively on the gap value we have fixed $\Delta=0.72$ meV which
%% is in a range consistent with neutron scattering data \scite{CavadiniHeigold}
%% and yields a good descriptition of the measured magnetization curves
%% \scite{NikuniOshikawa} as shown later on. Results for the superexchange
%% parameters from a fit with this gap value used as constraint are given in
%% table \ref{tab1}.
\begin{fulltable}
%%\caption{\label{tab1}Superexchange parameters from a fit of the triplet
%%  dispersion. The values in parenthesis are obtained by using
%%  the renormalized dispersion. Values for the scattering amplitude $v_0$ and
%%  the effective mass $m$ are also given.}
\caption{\label{tab1} Superexchange parameters, self-energies, scattering amplitude $v_0$ and
  effective mass $m$ for dispersion (\ref{eq2}) and (\ref{eq9}) (in brackets).}
%% \begin{ruledtabular}
{\footnotesize
\begin{tabular}{c|c|c|c|c|c|c|c|c|c}
$\!\!$$\Delta$ [meV]$\!\!$ &$\!\!$ $J$ [meV]$\!\!$&$\!\!$ $J_a$ [meV]$\!\!$&$\!\!$ $J_{a2c}$ [meV]$\!\!$&$\!\!$ $J_{abc}$ [meV]$\!\!$ &
$\!\!$$\Sigma_n$ [meV]$\!\!$&$\!\!$ $\Sigma_a$ [meV]$\!\!$&$\!\!$ $Z$$\!\!$ &$\!\!$ $v_0$ [meV]$\!\!$&$\!\!$ m [1/meV] \\
\hline\hline
%% 0.72 & 5.52 (4.76) &  -0.24 (-0.30) & -1.57 (-1.83) & 0.46 (0.57) & 0 (1.27) &
%% 0 (0.45)& 1 (0.91) & 7.91 (9.64) & 0.28 (0.29) 
%% 5.52214989 -0.23778465 3.13295892 0.91616544
%% 5.52257987 -0.23894084 3.13343436 0.91737987
%% 4.83601265 -0.30068947 3.68213706 1.16141042 
%% 4.83423381 -0.30225761 3.68378356 1.16369611
0.65 & 5.52 (4.83) & -0.24 (-0.30) & -1.57 (-1.84) & 0.46 (0.58) & 0 (1.24) & 0 (0.46) & 1 (0.91) & 8.01 (9.91)
& 0.25 (0.26) \\
\hline
%% 0.8 5.51903646 -0.22941008 3.12951578 0.90736913
%% 4.76725315 -0.28735243 3.64258347 1.13716933
0.8 & 5.52 (4.77) & -0.23 (-0.29) & -1.56 (-1.82) & 0.45 (0.57) & 0 (1.25)& 0 
(0.45)& 1 (0.91) & 7.77 (9.44) & 0.32 (0.32) \\
\hline
%% 1.0 5.51316651 -0.21360830 3.12302162 0.89077217
%% 4.77775298 -0.26611663 3.61501369 1.10657335
1.0 & 5.51 (4.78) & -0.21 (-0.27) & -1.56 (-1.81) & 0.45 (0.55) & 0 (1.20) & 0
(0.43) & 1 (0.92) & 7.39 (8.89) & 0.41 (0.42)
%% 5.52101779 -0.23474004 3.13170704 0.91296747
%% \hline\hline
%% $\Sigma_n$ [meV]& $\Sigma_a$ [meV]& $Z$ & $v_0$ [meV]& m [1/meV] \\
%% \hline
%% 0 (1.27) & 0 (0.45)& 1 (0.91) & 7.91 (9.64) & 0.28 (0.29) 
\end{tabular}
}
%% \end{ruledtabular}
\end{fulltable}

\begin{figure}[!htp]
\begin{center}
\includegraphics*[width=0.8\columnwidth]{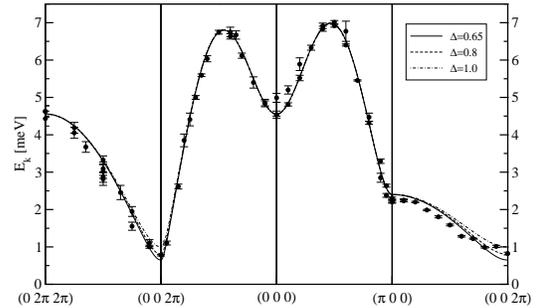}
\end{center}
\caption{Dispersion relation for TlCuCl$_3$ measured by INS (circles)
  \scite{CavadiniHeigold} in comparison to the theoretically calculated ones
  with fixed gap values $\Delta=0.65, 0.8, 1.0$~meV.}
\label{fig1}
\end{figure}

%%  \begin{table}
%% \caption{\label{tab1}Results for the superexchange parameters from fits with
%%   different constraint gaps. The values in parenthesis are obtained by using
%%   the renormalized dispersion. Values for scattering amplitude $v_0$ and
%%   triplet density $n_t$ are also given.}
%% \begin{ruledtabular}
%% \begin{tabular}{c|c|c|c|c}
%% $\Delta$ & $J$ & $J_a$ & $J_{a2c}$ & $J_{abc}$ \\
%% \hline
%% 0.72 & 5.521 (4.757) &  -0.235 (-0.295) & -1.566 (-1.825) & 0.456 (0.574)\\
%% %% 5.52101779 -0.23474004 3.13170704 0.91296747
%% \hline\hline
%% $\Sigma_n$ & $\Sigma_a$ & $Z$ & $v_0$&  \\
%% \hline
%% 0 (1.269) & 0 (0.453)& 1 (0.913) & 7.914 (9.642) & 
%% \end{tabular}
%% \end{ruledtabular}
%% \end{table}
%% \begin{table*}
%% \caption{\label{tab1}Results for the superexchange parameters from fits with
%%   different constraint gaps. The values in parenthesis are obtained by using
%%   the renormalized dispersion. Values for scattering amplitude $v_0$ and
%%   triplet density $n_t$ are also given.}
%% \begin{ruledtabular}
%% \begin{tabular}{c|c|c|c|c|c|c|c|c}
%% $\Delta$ & $J$ & $J_a$ & $J_{a2c}$ & $J_{abc}$ &$\Sigma_n$ & $\Sigma_a$ & $Z$ & $v_0$  \\
%% \hline
%% 0.5 & 5.53 (4.74)& -0.25 (-0.31) & -1.57 (-1.84)& 0.46 (0.59) & 0
%% (1.32)& 0 (0.47) & 1 (0.91) & 8.33 (10.23) \\
%% 0.7 &&&&&&&&\\
%% 0.8 & 5.52 (4.77) & -0.23 (-0.29) & -1.57 (-1.82) & 0.46 (0.57) & 0 (1.25)& 0
%% (0.45)& 1 (0.91) & 7.77 (9.44)\\
%% 1.0 & 5.51 (4.78) & -0.21 (-0.27) & -1.56 (-1.81) & 0.45 (0.55) & 0 (1.20) & 0
%% (0.43) & 1 (0.92) & 7.39 (8.89)
%% \end{tabular}
%% \end{ruledtabular}
%% \end{table*}
Next we discuss the renormalization of the dispersion (\ref{eq2}) due to the
hard-core constraint. Following \scite{KotovSushkov} this condition can be
taken into account by introducing an infinite on-site repulsion between the
bosons:
\begin{equation}
\label{eq4}
H_U = U\sum_{i,\alpha,\beta}
t^\dagger_{i\alpha}t^\dagger_{i\beta}t_{i\alpha}t_{i\beta}, \quad
U\rightarrow\infty\; .
\end{equation}
The corresponding scattering vertex
$\Gamma_{\alpha\beta,\alpha\beta}(\vec{K})$ where $\vec{K}=(\vec{k},\omega)$
is the total momentum and energy of the incoming particles can be calculated
exactly in the dilute limit by a summation of ladder diagrams. This yields
\begin{equation}
\label{eq5}
\Gamma(\vec{K}) =
-\left(\frac{1}{N}\sum_{\vec{q}} \frac{u_{\vec{q}}^2
    u_{\vec{k}-\vec{q}}^2}{\omega-\omega_{\vec{q}}-\omega_{\vec{k}-\vec{q}}}\right)^{-1} \; .
\end{equation}
%% independently of the spiral index of the scattered magnons. 
For magnetic fields $H\gtrsim H_g$ only particles near the band minimum at
$\vec{q}_0=(0,0,2\pi)$ are excited so that the energy and momentum dependent
$\Gamma(\vec{K})$ can be replaced by the constant $v_0=\Gamma(\vec{q}_0,0)$.
Because $u_{\vec{k}}^2$ is close to 1, $v_0$ is approximately given by the
magnon bandwidth $W$. However, it is known that the hard-core constraint
(\ref{eq4}) leads to a renormalisation of the Bogoliubov coefficients
(\ref{eq3}) and also to a renormalised triplet spectrum \scite{KotovSushkov}
%% given by
\begin{equation}
\label{eq9}  
\Omega_{\vec{k}0} =
Z_{\vec{k}}\sqrt{\left(A_{\vec{k}}+\Sigma_n(\vec{k},0)\right)^2-\left(B_{\vec{k}}+\Sigma_a\right)^2}
%% -\alpha g\mu_B H 
\; .
\end{equation}
Here $Z_{\vec{k}}^{-1}=1-\partial\Sigma_n/\partial\omega$ is the quasiparticle
residue, $\Sigma_n(\vec{k},\omega)$ the normal and $\Sigma_a$ the anomalous
self-energy. Values for these quantities and the renormalised superexchange
parameters
%% derived by a full self-consistent solution of the renormalization procedure 
are given in brackets in table \ref{tab1}. Although the superexchange
parameters are considerably renormalised, the shape of the dispersion
%% remains almost unchanged 
is only slightly changed because $\Sigma_n, \Sigma_a$ and $Z$ are almost
momentum independent here and, in addition, a fit to the measured dispersion
is performed. The most important consequence is a renormalisation of the
scattering amplitude $v_0$ by more than 20\%. The dispersions for all 3
gap-values are shown in Fig.~\ref{fig1} in comparison to the INS-data.
In all cases the experimental data are well described and differences between
the 3 fits are only visible close to $\vec{q}_0$.
%% However, we will show that these small differences have
%% drastic consequences for the boson density at low temperatures.
\section{HFP approximation}
\label{HFPA}
\begin{figure}[!htp]
\includegraphics*[width=0.99\columnwidth]{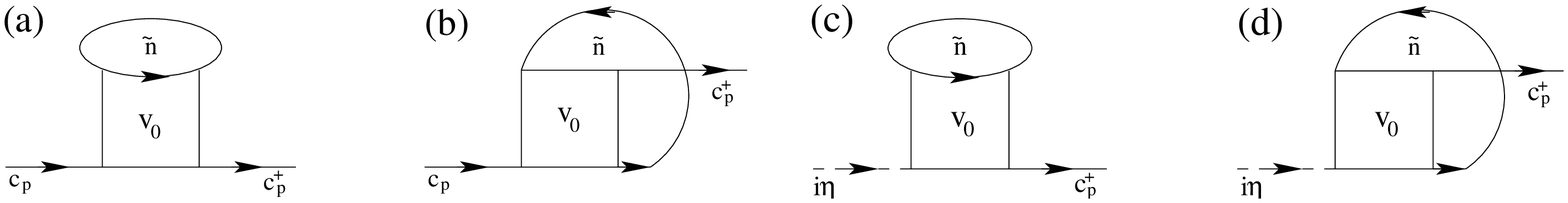}
\caption{(a), (b): Hartree and Fock diagram. (c), (d): Additional one-loop
  diagrams due to the terms cubic in $c_{\vec{k}}^{(\dagger)}$.}
\label{fig2}
\end{figure}

As we are interested in temperatures $T<\Delta\ll v_0$ we can treat $v_0$ as 
temperature independent and it is also sufficient to take only the lowest
triplet mode ($\alpha=+$) into account. For simplicity
%% we will from now on denote 
%% $t_{\vec{k}} \equiv \tilde{t}_{\vec{k},+}$,
we define $\epsilon_{\vec{k}}\equiv\Omega_{\vec{k}0}-\Delta$. The Hamiltonian
for the lowest triplet mode is then given by
\begin{equation}
\label{eq6}  
H = \sum_{\vec{k}} \left(\epsilon_{\vec{k}}-\mu_0\right)t_{\vec{k}}^\dagger t_{\vec{k}}
+\frac{v_0}{2}\sum_{\vec{k},\vec{k}',\vec{q}} t_{\vec{k}+\vec{q}}^\dagger t_{\vec{k}'-\vec{q}}^\dagger
t_{\vec{k}} t_{\vec{k}'} 
\end{equation}
where $\mu_0 = g\mu_B (H-H_g)$. To allow for the symmetry breaking in the
condensed phase we introduce new operators $t_{\vec{k}} = c_{\vec{k}} +
i\delta_{\vec{k},\vec{q}_0}\eta$ where $\eta$ is a real number and
$n_0=\eta^2$ the condensate density. To diagonalise (\ref{eq6}) we have to
deal with the terms which are cubic or quartic in the new operators
$c_{\vec{k}}$. The simplest way to treat these terms is the one-loop
approximation which yields the Hartree and the Fock diagram for the quartic
term (see Fig.~\ref{fig2}).
Therefore this approximation is often called Hartree-Fock-Popov approximation
(HFPA). As a result we find $H = H_c + H_{\text{lin}} + H_{\text{bilin}}$
where 
\begin{eqnarray}
\label{eq6b}  
H_c &=& \frac{v_0}{2}n_0^2-\mu_0n_0 \nonumber\\
H_{\text{lin}} &=&
\im(2\tilde{n}v_0\eta+v_0n_0\eta-\mu_0\eta)(c_{\vec{q}_0}^\dagger
-c_{\vec{q}_0}) \\
H_{\text{bilin}} &=& \sum_{\vec{k}} \left\{ \mathcal{A}_{\vec{k}} c_{\vec{k}}^\dagger c_{\vec{k}} - \frac{\Sigma_{12}}{2}\left(c_{\vec{k}}^\dagger
    c_{-{\vec{k}}}^\dagger + h.c.\right) \right\} \nonumber
\end{eqnarray}
with $\tilde{n}$ being the density of non-condensed magnons. Here
$\mathcal{A}_{\vec{k}} = \epsilon_{\vec{k}} - \mu_0+\Sigma_{11}$ with the
normal self-energy $\Sigma_{11} = 2v_0\tilde{n}+2v_0n_0$ and the anomalous
self-energy $\Sigma_{12} = v_0n_0$. By a standard Bogoliubov transform
$H_{\text{bilin}}$ can be diagonalised and
$E_{\vec{k}}=(\mathcal{A}_{\vec{k}}^2-\Sigma_{12}^2)^{1/2}$ is the
quasiparticle spectrum. In addition $H_{\text{lin}}$ has to vanish.

It is important to note that the HFPA is parametrically justified in the
present case: The magnon density $n=\tilde{n}+n_0$ is small and diagrams taken
into account in the HFPA are the leading diagrams in a systematic expansion in
$n$.  However, when $T\rightarrow T_c$ from below, where $T_c$ is the critical
temperature of the Bose gas, it is well known that the condensate density $n_0$
in the HFPA does not vanish.  Therefore the total density jumps and the HFPA
falsely predicts a first order phase transition \scite{GriffinRep}. This is due
to the fact that any perturbative approach fails at the critical point where
the scale for fluctuations diverges. By estimating when the next-leading
diagrams become as important as the diagrams taken into account one finds
\scite{GriffinRep} that the HFPA is justified when
\begin{equation}
\label{eq7}  
\left| T_c-T \right| \gg an^{1/3} T_c \; . 
\end{equation}
As $an^{1/3}$ is our small parameter, the temperature region around the
critical point where this perturbative approach is not justified is usually
very small.

In many papers about BEC in TlCuCl$_3$ much interest has focused onto the fact
that $H_c(T)-H_g\sim T^\phi$ with $\phi\sim 2$ \scite{NikuniOshikawa} where
$H_c$ is the critical field where BEC occurs. This has lead to a discussion
why the ``critical exponent'' $\phi$ deviates from $3/2$ which is the result
within the HFPA provided that the magnon dispersion is quadratic.  In the same
way one can also consider $n_c(T) \sim T^\phi$ and finds from the experimental
data \scite{NikuniOshikawa} that this ``power law'' yields a good fit with a
similar exponent $\phi\sim 2$. In HFPA one finds $n_c(T)\sim T^{3/2}$.
Therefore it has been speculated \scite{WesselOlshanii,NikuniOshikawa} that the
HFPA fails, i.e., that the ``critical exponent'' is renormalised due to
quantum fluctuations.

\begin{figure}[!htp]
\begin{center}
\includegraphics*[width=0.8\columnwidth]{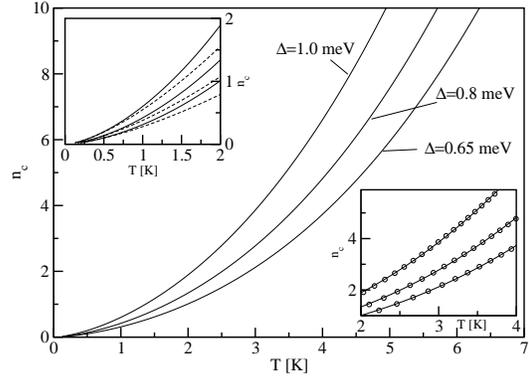}
\end{center}
\caption{$n_c(T)$ within HFPA with dispersions from
  Fig.~\ref{fig1}. Upper inset: For $T<1$ K the dispersions are almost
  quadratic and $n_c\sim T^{3/2}$ (dashed lines). Lower inset: $n_c(T)$ is
  well fitted by a ``power law'' (symbols) with exponents $1.92$ ($\Delta=0.65$
  meV), $1.88$ ($\Delta=0.8$ meV) and $1.82$ ($\Delta=1.0$ meV) for $2<T<4$ K.}
\label{fig3}
\end{figure}

First, it is worth to note that we are dealing with quantities which are not
dimensionless. Therefore the exponent is dictated by dimensional analysis. If
the dispersion is given by $\epsilon\sim k^\alpha$ we will find in
$d$-dimensions that $n_c(T)\sim T^{d/\alpha}$. Of course, the proportionality
factor is not necessarily temperature independent as in the HFPA. But it can
only depend on the dimensionless quantity $aT^{1/\alpha}$ (or equivalently
$an^{1/3}$ which is small for a dilute gas) where the scattering length $a$ is
the only length scale in our problem. Corrections of this kind have indeed
been found for interacting Bose gases \scite{BaymBlaizot}. As a consequence
there is no longer a simple power law. However, for the BEC of magnons there
is also another effect which makes a simple power law invalid even within
HFPA: The dispersion is certainly quadratic for small excitation energies but
the range of validity of the quadratic approximation might be rather small. As
we pointed out \scite{SirkerWeisse,SirkerWeisse2} this is indeed the case for
TlCuCl$_3$ where the quadratic approximation is only justified for $T<1$ K
which is well below the experimental temperature range \scite{NikuniOshikawa}.
To see this we show in Fig.~\ref{fig3} $n_c(T)$ calculated within the HFPA
using the triplet dispersions from Fig.~\ref{fig1}.
Whereas for $T<1$ K we indeed see that $n_c\sim T^{3/2}$ there is no power law
at higher temperatures. However, as shown in the lower inset of
Fig.~\ref{fig3} it is still possible to obtain an excellent fit by a ``power
law'' even at higher temperatures provided that the considered temperature
range is sufficiently small. For the temperature range in
\scite{NikuniOshikawa} we find an exponent $\phi\sim 1.8-1.9$ which is in good
agreement with experiment and a recent work \scite{MisguichOshikawa} where also
a realistic dispersion has been used. We should not take this agreement too
serious because the HFPA is not justified at the critical point.  In addition,
we will argue in the next section that anisotropies play a crucial role and
that actually no phase transition occurs.
%% There will be still pronounced minima $n^*$ in the magnetisation curves,
%% however, $n^*(T)$ will depend on the anisotropy even if the magnon dispersion
%% is quadratic. 
What Fig.~\ref{fig3} nevertheless does show is that in the experimental
temperature range $n_c(T)$ does depend on microscopic details and no universal
power law exists.
\section{Spin-phonon and spin-orbit coupling}
\label{secSO}
\begin{figure}[!htp]
\includegraphics*[width=0.99\columnwidth]{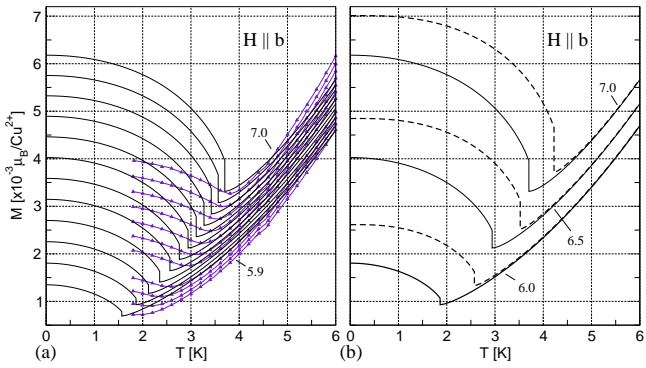}
\caption{(a) Experimental magnetisation curves (symbols) for
  $H=5.9,6.0,\cdots,7$~T from Ref.~\scite{NikuniOshikawa} compared to the
  theoretically calculated (solid lines) with $\Delta = 0.67$~meV, $v_0 =
  25$~meV. (b) Theoretical magnetisation curves as in (a) (solid lines) and
  with an additional exchange anisotropy $\tilde{\gamma}=0.01$ meV (dashed
  lines).}
\label{fig4}
\end{figure}

In \scite{SirkerWeisse} we have shown that the magnetisation $M(T)$ calculated
%% using the HFPA 
with the dispersion (\ref{eq9}) and $v_0\sim 10$ meV as obtained by the BO
formalism does not fit the experimental data.  However, our calculation of
$v_0$ in section \ref{secBO} does only include magnon-magnon scattering. From
Raman spectroscopy \scite{ChoiGuentherodt} and sound attenuation experiments
\scite{ShermanLemmens} it is known that phonons are important.
%% and do couple to the spin system. 
Because optical phonon modes exist at energies comparable with the energy
scale of the magnetic excitations \scite{ChoiGuentherodt} it is not clear if
%% the renormalisation of $v_0$ due to 
the effect of magnon-phonon scattering can be calculated perturbatively. We
have therefore used $v_0$ as a fit parameter and obtained best agreement with
experiment for $v_0\approx 25$~meV as shown in Fig.~\ref{fig4}a.
A renormalisation of $v_0$ by a factor $2-3$ might be caused by a reduction of
the bare magnon bandwidth due to polaronic effects. But even with $v_0=25$~meV
we can obtain good agreement with experiment only for $T>T_c$. At low-T our
theory still overestimates $M(T)$ by 50\%. $an^{1/3}\sim 0.1$ so that HFPA
should be justified according to (\ref{eq7}) apart from a region $\sim\pm 0.5$
K around the critical point. The approximation therefore cannot explain the
failure of our theory at low-T.

In any real magnetic system there is some kind of anisotropy which reduces the
symmetry. In a system without magnetic field and anisotropies we have the
usual $SU(2)$ symmetry. By a magnetic field this symmetry is reduced to $U(1)$
around the magnetic field axis. Spontaneous breaking of $U(1)$ occurs at the
BEC transition and is responsible for a gapless Goldstone mode in the phase
with $n_0\neq 0$. Any kind of anisotropy will in general break $U(1)$
explicitly so that there is no longer a Goldstone mode. However, depending on
the type of anisotropy there might be still a $\mathcal{Z}_2$ symmetry
(changing the sign of the triplet operator) so that a transition between a
phase with and without condensed magnons is still possible. Finally, if the
anisotropy breaks $\mathcal{Z}_2$ no phase transition will occur and for a
small symmetry breaking anisotropy we expect the phase transition to be
smeared out to a crossover.

\begin{figure}[!htp]
\includegraphics*[width=0.99\columnwidth]{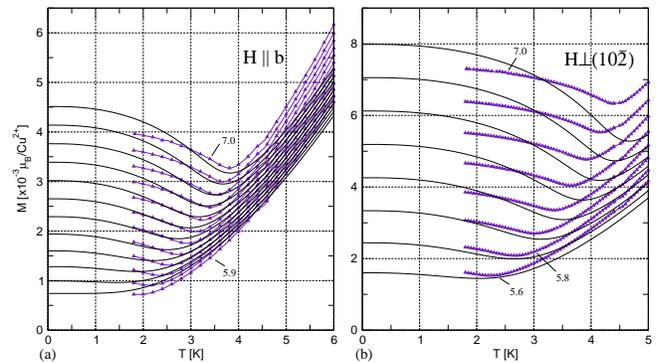}
\caption{(a): Experimental magnetisation curves as in Fig.~\ref{fig4}a and
  theoretically calculated with $g=2.06$ \scite{OosawaIshii}, $\Delta=0.72$
  meV, $v_0=27$~meV and $\gamma=10^{-3}$~meV. (b): As in (a) with $g=2.26$
  \scite{OosawaIshii}.}
\label{fig5}
\end{figure}

Here we want to discuss two kinds of anisotropy.
%% with and without breaking $\mathcal{Z}_2$. 
First, we want to consider an exchange anisotropy (EA) within a dimer, i.e.,
we consider $J\rightarrow \{J^x,J^y,J^z\}$. Using triplet operators,
performing the Bogoliubov transformation with parameters as in Eq.~(\ref{eq3})
and considering only the lowest triplet mode we find
\begin{equation}
\label{SO1}  
H_{1,\text{pert}}=\tilde{\gamma}(t_{\vec{k}}t_{-\vec{k}}+t^\dagger_{\vec{k}}t^\dagger_{-\vec{k}})
\end{equation}
as perturbation to (\ref{eq6}) where $\tilde{\gamma}\propto J_x-J_y$. This
kind of perturbation can also originate from a ``single-ion anisotropy'' for
the triplets $\sim D(S^z_1+S^z_2)^2+E[(S^x_1+S^x_2)^2-(S^y_1+S^y_2)^2]$ where
$\vec{S}_{1,2}$ denote the spins within one dimer. 
%% Clearly $\mathcal{Z}_2$ symmetry is still there and we therefore still expect
%% a phase transition. 
On the other hand consider a Dzyaloshinsky-Moriya anisotropy (DMA)
$\sim\vec{D}\cdot\left(\vec{S}_1\times\vec{S}_2\right)$ within the dimer.
Transforming this type of interaction into triplet operators we find
\begin{equation}
\label{SO2}  
H_{2,\text{pert}}=\im\gamma (t_{\vec{q_0}} - t^\dagger_{\vec{q_0}}) 
\end{equation}
where the wave vector $\vec{q}_0$ depends on how the DM-vector $\vec{D}$
varies in space \scite{SirkerWeisse}. In particular we want to consider the
case where $\vec{q}_0$ corresponds to the minimum of the triplet dispersion.
%% Because (\ref{SO2}) is linear in the triplet operators the $\mathcal{Z}_2$
%% symmetry is broken and no phase transition will occur.
Both types of anisotropy yield additional contributions to the Hamiltonian
(\ref{eq6b}) and results for the quasiparticle spectra and the additional
constraints due to the linear term are given in table \ref{tab2}.
\begin{fulltable}[!htp]
\caption{\label{tab2} Comparison between exchange and DM-type anisotropy}
{\small
 \begin{tabular}{c|c|c}
& exchange anisotropy ($\tilde{\gamma}>0$) & DM-type anisotropy ($\gamma\neq 0$)\\[0.1cm]
\hline\hline
& $-\mu_0\eta -2\tilde{\gamma}\eta+ \eta\left(2v_0\tilde{n}+v_0n_0\right)= 0$
& $-\mu_0\eta -\gamma + \eta\left(2v_0\tilde{n}+v_0n_0\right)= 0$ \\
$H_{\text{lin}} = 0$ &Two cases: $\eta=0$ (non-condensed phase) & always condensed magnons,
$\eta\neq 0$ \\
& and $\eta\neq 0$ (condensed phase) & \\[0.1cm]
\hline
&&\\[-0.3cm]
Phase transition& $n_c=\frac{\mu_0}{2v_0}+\frac{\tilde{\gamma}}{v_0}$ & no
phase transition\\[0.1cm]
\hline
&&\\[-0.3cm]
non-condensed phase & $E_{\vec{k}} = \sqrt{(\epsilon_{\vec{k}}-\mu_0+2\tilde{n}v_0)^2-4\tilde{\gamma}^2}$ &
------------------\\[0.1cm]
\hline
&&\\[-0.3cm]
condensed phase & $E_{\vec{k}} = \sqrt{(\epsilon_{\vec{k}}+2\tilde{\gamma})^2+2(\epsilon_{\vec{k}}+4\tilde{\gamma})v_0n_0-4\tilde{\gamma}^2}$ & $E_{\vec{k}} =\sqrt{\left(\epsilon_{\vec{k}}+\frac{|\gamma|}{\sqrt{n_0}}\right)^2+2\left(\epsilon_{\vec{k}}+\frac{|\gamma|}{\sqrt{n_0}}\right)v_0n_0}$\\
\end{tabular}
}
\end{fulltable}
When compared to the case without anisotropy we see that the EA yields a small
shift $\tilde{\gamma}/v_0$ in $n_c$ and also a small quasiparticle gap
$\Delta_{qp}=\sqrt{8\tilde{\gamma}v_0n_0}$ in the condensed phase. At the
critical point, $\Delta_{qp}$ is zero. Fig.~\ref{fig4}b shows that apart from
the shift in $n_c$ the shape of the magnetisation curves is basically
unaffected. On the other hand a DMA has a dramatic effect even if it is tiny
because it smears out the phase transition to a crossover. As already shown in
\scite{SirkerWeisse} it is possible to obtain excellent agreement with the
measured magnetisation curves for $H\parallel b$ when a DMA
$\gamma=10^{-3}$~meV is included (see Fig.~\ref{fig5}a). In an ideal
TlCuCl$_3$ crystal the centre of each dimer is an inversion centre.
$\gamma\neq 0$ therefore requires small lattice distortions. If these
distortions are of such kind that $\vec{D}$ is oriented along a specific
direction throughout the crystal it would be possible to restore $U(1)$
symmetry by applying the magnetic field along the same axis. In this
configuration a sharp phase transition would still occur.
We therefore compare here also with experimental results for $H\perp
(10\bar{2})$ \scite{OosawaIshii} in Fig.~\ref{fig5}b. As for $H\parallel b$ the
magnetisation curves show only a smooth increase at temperatures below the
minima and no sharp phase transition. Therefore a component of $\vec{D}\perp
H$ seems to exist also for this configuration. This could be possibly
explained by small lattice distortions leading to domains with different
orientation of $\vec{D}$. In this case a component $\vec{D}\perp H$ could
exist for each field configuration. The theoretically calculated $M(T)$ show
reasonable agreement with experiment also for $H\perp (10\bar{2})$ if we
change the $g$-factor from $2.06$ to $2.26$ according to ESR
\scite{OosawaIshii,GlazkovSmirnov}.
%% When calculating the magnetisation curves
%% theoretically with all parameters as for $H\parallel b$ except for the $g$
%% factor which has been measured to be $g\approx 2.25$ for $H\perp (10\bar{2})$
%% instead of $g\approx 2.06$ for $H\parallel b$
%% \scite{OosawaIshii,GlazkovSmirnov} we indeed have reasonable agreement with
%% experiment also for $H\perp (10\bar{2})$ (see Fig.~\ref{fig5}b). 
This shows that the variation in $\gamma$ with field direction and also
additional crystal field anisotropies seems to be relatively minor. Although
we could certainly improve agreement with experiment by including such effects
we have not done so because the number of fit-parameters would be too large.

\begin{figure}[!htp]
\includegraphics*[width=0.99\columnwidth]{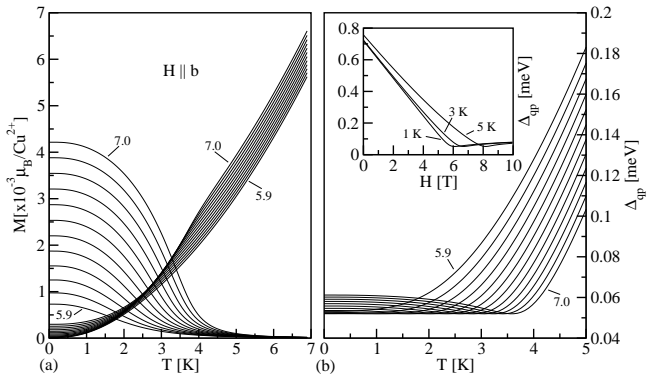}
\caption{(a) Condensed and non-condensed magnon densities. (b) $\Delta_{qp}$
  as function of $T$ for different $H$ and as function of $H$ for $T=1$ K, $3$
  K and $5$ K (inset). The parameters are as in Fig.~\ref{fig5}a.}
\label{fig6}
\end{figure}

In Fig.~\ref{fig6}a we show the condensed and the non-condensed density
separately for the same parameters as in  Fig.~\ref{fig5}a. 
%% Due to the small exchange anisotropy we see that the condensed density does
%% not drop down sharply. 
The transition from the condensed to the non-condensed phase is smeared to a
crossover in a region $\sim 1-2$ K around the former transition point.
Analytically we find that the temperature region $\Delta T$ where the
densities with $\gamma\neq 0$ deviate significantly from those without such an
anisotropy is given by $\Delta T / T \sim (\gamma/v_0)^{2/3}/n$.

%% \begin{equation}
%% \label{SO3}  
%% \frac{\Delta T}{T}\sim \frac{1}{3n}\left(\frac{\gamma}{v_0}\right)^{2/3}\; .
%% \end{equation} 
%% For the parameters used here this is in agreement with the findings from
%% Fig.~\ref{fig6}a. 
As the temperature range where quantum fluctuations are important is much
smaller according to Eq.~(\ref{eq7}) the HFPA is parametrically justified here
even in the crossover region. This does not mean that corrections to the HFPA
do not exist but these corrections can be calculated everywhere in a
perturbative manner.

Finally, we show in Fig.~\ref{fig6}b the quasiparticle gap $\Delta_{qp}$. As a
function of temperature $\Delta_{qp}$ has a minimum basically at the same
point where also the magnetisation curves in Fig.~\ref{fig5}a have their
minima. The theoretically calculated $\Delta_{qp}$ as a function of magnetic
field at fixed temperature does qualitatively agree with recent ESR
measurements \scite{GlazkovSmirnov}. Quantitatively the gap predicted by our
theory is about a factor 2 smaller than the one measured by ESR for
$H\parallel b$ at $H=9$ T. Two explanations are possible: First, within our
theory it is also possible to obtain reasonable agreement with the measured
magnetisation curves for $\gamma\sim 5\cdot 10^{-3}$ meV by changing $v_0$ and
the excitation gap $\Delta$ accordingly. In this case $\Delta_{qp}\sim 0.2$
meV in agreement with ESR. Second and more important, it is not clear if the
gap measured in ESR is the pure quasiparticle gap. Within our theory we would
expect the gap measured in ESR at fields above the ``critical field'' (minimum
in the gap function) to be given by a combination of one- and two-magnon
excitations.  In this case the ESR gap would be a combination of the real
quasiparticle gap $\Delta_{qp}$ and $2\Delta_{qp}$ \scite{SirkerSushkov}. To
avoid these ambiguities INS measurements of the gap are desirable.
\section{Specific Heat}
\label{specHeat}
\begin{figure}[!htp]
\includegraphics*[width=0.99\columnwidth]{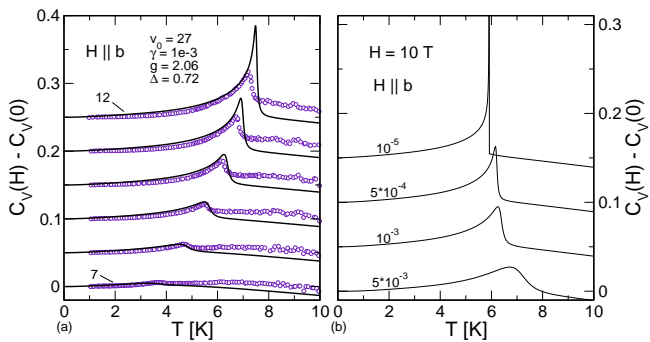}
\caption{(a) Measured specific heat \scite{OosawaKatori} (per copper ion) for $H=7,8,\cdots,12$ T
  (circles) and theoretically calculated (lines). (b) Calculated specific heat
  for DMA $\gamma=5\cdot 10^{-3},\cdots,10^{-5}$. In both graphs the
  subsequent curves are shifted by $0.05$.}
\label{fig7}
\end{figure}

To calculate the specific heat $C_V(H)$ we need the energy $E_+(H)$ of the
lowest triplet component ($\alpha=+$) in the presence of a DMA $\gamma$ and a
magnetic field $H$. From Eq.~\ref{eq6b} and table \ref{tab2} one easily finds
\begin{eqnarray}
\label{SH1}  
E_+(H) &\!\!\! =\!\!\! & \mu_0n_0 -
(\tilde{n}+n_0)v_0-\frac{3}{2}v_0n_0^2-4v_0\tilde{n}n_0 \\
&\!\!\!\!\!\!\!\!\!\!\!\!\!\! -&\!\!\!\!\!\!\!\!\! v_0\tilde{n}^2 + \frac{1}{2N}\sum_{\vec{k}}(E_{\vec{k}}-\epsilon_{\vec{k}})+\frac{1}{N}\sum_{\vec{k}} E_{\vec{k}} \tilde{n}_{\vec{k}} +\frac{\mu_0}{2}\nonumber
\end{eqnarray}
where $\gamma$ only enters by the condition for the vanishing of the linear
term (see table \ref{tab2}). As we want to calculate the specific heat for
temperatures $T\sim\Delta$ we also have to take into account the contribution
of the triplet mode $\alpha = 0$. At $H=0$ all modes are of the same
importance and the DMA yields only tiny corrections which we can ignore.
Therefore the energy $E(0)$ at zero field of each triplet mode $\alpha =
+,-,0$ is simply given by
\begin{eqnarray}
\label{SH2}  
E(0) &=& -2v_0\tilde{n}^2
+\frac{1}{N}\sum_{\vec{k}}(\Omega_{\vec{k}0}+4v_0\tilde{n})\tilde{n}^0_{\vec{k}} \\
\tilde{n}^0_{\vec{k}} &=& 1/(\exp[\beta (\Omega_{\vec{k}0}+4v_0\tilde{n})]-1)
\; .\nonumber
\end{eqnarray}
As phonons yield a large contribution to the specific heat which we do not
want to consider here, we only compare theoretical and experimental results
for $C_V(H)-C_V(0)$. In this quantity most phonon contributions should be
eliminated. As the $\alpha=0$ mode is basically unaffected by the magnetic
field we find
\begin{equation}
\label{SH3}
C_V(H)-C_V(0) = \frac{\partial \left[E_+(H)-2E(0)\right]}{\partial T} \; .
\end{equation}
The results are shown in Fig.~\ref{fig7}a in comparison to the experimental
data \scite{OosawaKatori}. 
The agreement is good. The overestimation of the peak heights and slight
underestimation of the widths indicates that $\gamma$ seems to be slightly
larger than assumed here. In addition we also see some deviations at higher
temperatures particularly at higher fields. Here we should remember that at
such energies we have to use in principle the energy and momentum dependent
scattering amplitude $\Gamma(\vec{K})$ instead of the constant $v_0$. The
dependence of peak height and width on $\gamma$ is shown in Fig.~\ref{fig7}b.
By increasing $\gamma$ the peak gets smaller and broader. Note that without
anisotropy there would be a singularity in $C_V(H)-C_V(0)$ at the critical
point due to the failure of the HFPA.
\section{Conclusions}
To summarise, we have shown how to incorporate the hard-core constraint into
the BO formalism if the magnons are dilute. For TlCuCl$_3$ we have found that
the dispersion is only slightly modified when compared to
\scite{MatsumotoNormandPRL} where the constraint has been completely ignored.
This is due to the following facts: (a) The quasiparticle residue and the
self-energies turn out to be almost momentum independent and (b) a fit to the
measured dispersion is performed. However, even in this case the correct
treatment of the constraint is important when calculating the magnon-magnon
scattering amplitude $v_0$. We have solved the hard-core boson model using the
HFPA and have shown that this approximation is valid apart from a small region
around the critical point. Even if TlCuCl$_3$ is assumed to be a system
without anisotropies no power law $n_c~\sim T^\phi$ can be expected in the
experimental temperature range \scite{NikuniOshikawa} because the quadratic
approximation for the dispersion works only for $T<1$ K. Next, we have
discussed how tiny EA or DMA influence BEC. An EA yields only a small shift in
$n_c$ and a small quasiparticle gap in the condensed phase but leaves the
shape of the magnetisation curves otherwise unchanged. On the other hand a DMA
has a dramatic effect and smears out the phase transition to a crossover. The
different effects of these anisotropies can be understood by the symmetries of
the system. With a DMA $\gamma=10^{-3}$ meV we have achieved good agreement
with experimental data for $M(T)$ and $C_V(H,T)$ if $H\parallel b$. When
rescaled by the $g$-factor the agreement with $M(T)$ for $H\perp (10\bar{2})$
was also reasonable. This shows that a component of the DM-vector
$\vec{D}\perp H$ seems to exist for each field configuration and that
additional crystal field anisotropies are relatively minor. Within our theory
we expect a quasiparticle gap $\Delta_{qp}\sim 0.1$ meV for $H\parallel b,\,
H\sim 10$ T which might be tested in the future.  \acknowledgment The authors
thank Ch.~R\"uegg for providing the INS-data and acknowledge support by the
ARC. JS acknowledges valuable discussions with M.~Oshikawa and I.~Affleck and
support by the DFG.  

\end{document}